# A novel method to calculate the electric field using solid angles

Fulin Zuo

Department of Physics

University of Miami

Coral Gables, FL 33124

## Abstract
A new method to calculate the electric field inside a spherical shell with surface charge $\sigma = \sigma_o \cos\theta$ in terms of solid angle is presented. The integral can be readily carried out without invoking special functions typically used for this classical problem. For a flat surface of uniform charge density, the electric field normal to the surface is shown to be proportional to the solid angle subtended by the surface only.

## Introduction

The electric field inside a uniformly charged spherical shell is a standard example discussed in the introductory university physics. The field inside is zero and on the outside it behaves like a point charge at the center. This can be traced back the original Newton's shell theorem for gravitational force [1] and it is usually proven with the use of Gauss's law and spherical symmetry of the electric field [2, 3]. In fact, students are usually very surprised to find out the contributions from the small dome close to a point is cancelled out exactly by the larger dome farther away if one separates the spherical shell into two parts by a plane containing the point and perpendicular to the line connecting the point and the center of sphere.

Solid angle is a key concept to the proof of Gauss's law. The notation is often introduced in calculus based college physics textbooks and but usually not much more beyond the definition. It can be used to easily prove the exact cancellation of electric fields from opposing spherical segments [1,2].

Field calculations for more complicated charge distributions, such as a charged ring or a disk are usually limited to points on the symmetry axis. General solutions are typically reserved as boundary value problems in more advanced courses [4, 5]. However, new formulations or treatments are still of current interest [6-13].

One classical E&M problem is the field inside a spherical shell with a charge density varying as $\sigma = \sigma_o \cos\theta$. It describes of charge distribution of several systems such as the bound surface charge density of a uniformly polarized sphere or the induced surface charge density of a spherical conductor inside a uniform electric field. In the case of uniformly polarized sphere $\sigma_o = p$, $p$ is the dipole moment per unit volume. The calculation of the field is a standard example of boundary value problem, easily found in any intermediate electromagnetism books [4]. It should also be noted for this charge density there exists another clever method to show the field is uniform by superposing two oppositely charged spheres of charge Q separated by vanishingly small distance d but with a finite diploe moment of Qd.

## Results

Here we demonstrate that field inside a sphere due to this charge density $\sigma = \sigma_o \cos\theta$ is constant by a simple integration with the use of solid angle concept.

Consider an arbitrary point P inside a sphere as shown in Figure 1, where the vertical axis corresponds to z-axis or $\theta = 0$. From point P construct two intersecting infinitesimal cones with P at the vortex. The two infinitesimal cones of same solid angle $d\Omega$ intersect with the sphere with an infinitesimal area of dA and dA' respectively. If dA is defined by a spherical polar angle Θ and azimuthal angle Ø, dA' is then at a polar angle of γ=π-2α-Θ. The line connecting dA and dA' is at angle of β=α+Θ with respect to z-axis. The solid angle $d\Omega$ is related to the area dA by $d\Omega = \frac{dA \cdot \hat{n}}{r^2} = \frac{dA}{r^2}\cos\alpha = \frac{dA'\cdot\widehat{n'}}{r'^2} = \frac{dA'}{r'^2}\cos\alpha$, where r and r' are the distances from P to dA and dA' respectively.

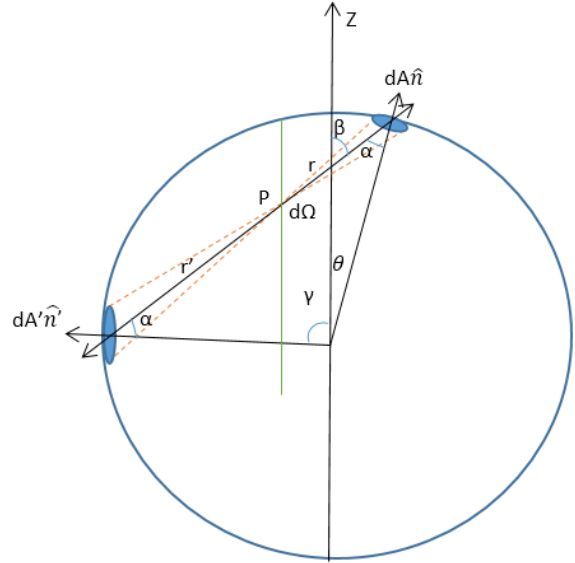

Figure 1 Electric field due to two intersecting infinitesimal cones on a spherical surface

The electric field from dA and dA' are $dE = \frac{k\sigma(\theta)dA}{r^2} = \frac{k\sigma(\theta)d\Omega}{\cos\alpha}$ and $dE' = \frac{k\sigma(\gamma)dA'}{r'^2} = \frac{k\sigma(\gamma)d\Omega}{\cos\alpha}$.

If the charge density is the same, the two contributions balance out completely, thus zero net field anywhere inside a uniformly charged spherical shell.

If the charge density is in the form $\sigma = \sigma_0 \cos\theta$, the net electric field contribution

$dE_n = dE - dE' = \frac{kd\Omega}{\cos\alpha}[\sigma(\theta) - \sigma(\gamma)]$, along the $-\hat{r}$ direction ($\hat{r}$ is from P to dA). Because it is only dependent on the polar angle Θ, independent of Ø, the horizontal component cancels out and leaving only the vertical component:

$$dE_z = -dE_n \cos\beta = -\frac{kd\Omega}{\cos\alpha}\sigma_0[\cos\theta - \cos\gamma]\cos\beta = -\frac{kd\Omega}{\cos\alpha}\sigma_0\left[-2\sin\frac{\theta-\gamma}{2}\sin\frac{\theta+\gamma}{2}\right]\cos\beta$$

$$= -\frac{kd\Omega}{\cos\alpha}\sigma_0\, 2\cos(\theta+\alpha)\cos\alpha\cos\beta = -2k\sigma_0(\cos\beta)^2\, d\Omega = -2k\sigma_0(\cos\beta)^2\, 2\pi\sin\beta\, d\beta$$

The total electric is then given by

$$E = E_z = -\int_0^{\frac{\pi}{2}} 2k\sigma_0(\cos\beta)^2\, 2\pi\sin\beta\, d\beta = -\frac{4\pi k\sigma_0}{3}$$

The integral does not require special functions such as the Legendre polynomials, except the notation for solid angle in spherical coordinates.

In general, one can express the infinitesimal field contribution from any infinitesimal charge on the surface and integrate over to get the same result, but in reality the expression is usually too complicated to contemplate.

Another interesting geometry where solid angle can be used directly to calculate the electric field is that of a flat surface of uniform charge density. Assuming a uniform charge density σ is on the xy plane, as in Figure 2(a), the electric field component in the z-axis direction anywhere due to an infinitesimal charge $dq = \sigma dA$ is given by

$$dE_z = \frac{k\sigma dA}{r^2} \hat{r} \cdot \hat{z} = \frac{k\sigma d\vec{A}\cdot\hat{r}}{r^2} = k\sigma d\Omega.$$

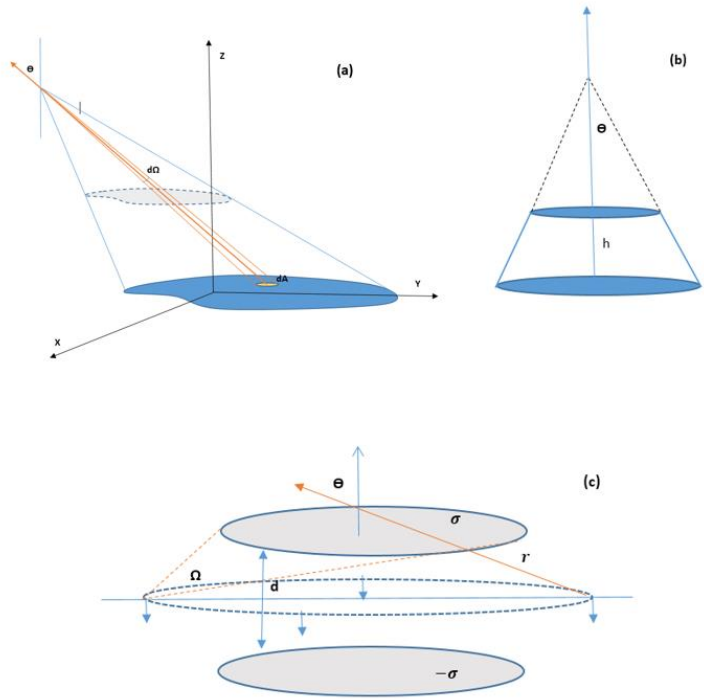

The total field in the z-axis direction is thus $E_z = k\sigma\Omega$ if the charge density is constant, here Ω is the solid angle subtended by the surface from the point. The result is interesting that it only depends on the solid angle, independent of the distance from the plane.

Figure 2 (a) dE from an infinitesimal surface charge; (b) E at the tip of a truncated cone; (c) E on the plane bisecting the plates

For example, the surface bounded by dashed line in Figure 2(a) would contribute exactly the same field in the z-axis direction.

The result is especially useful if the point is on the symmetry axis, such that the axial component is the total field. A familiar example is that the field due to a uniformly charged circular dish, given by

$E = k\sigma\Omega = k\sigma 2\pi(1 - \cos\theta)$. A trivial example is when $\Omega = 2\pi$ for an infinitely large sheet with $E = 2\pi k\sigma$.

A fun example is that of the electric field at the tip of a uniformed charged cone or truncated cone. Here the contribution from each layer of thickness dz contributing exactly the same $dE = k\rho dz\Omega$, and the total field $E = kh\rho\Omega = kh\rho 2\pi(1 - \cos\theta)$, where ρ is the uniform volume charge density and h is the height of the truncated or full cone, as in Figure 2(b).

The result can also be used to calculate the edge field for a parallel plate capacitor[13]. For example, the electric field on the plane bisecting the plates separated by a distance d is simply given by $E = 2k\sigma\Omega$, assuming σ is constant. For far away points, $\Omega \approx \frac{A\hat{n}\cdot\hat{r}}{r^2} = \frac{A}{r^2}\cos\theta$, the result gives the well known dipole electric field

$$\text{E} = \frac{A\hat{n}\cdot\hat{r}}{r^2} = \frac{2k\sigma A}{r^2}\cos\theta = \frac{kP}{r^3}, \text{ with } P = \sigma Ad,\text{ as shown in Figure 2(c).}$$

## Conclusion

In summary, we have presented a new way with the aid of solid angle concept to calculate the electric field inside a spherical shell of charge density $\sigma = \sigma_o \cos\theta$. To our knowledge, this is a new method to solve this problem and the present approach is not available or published in the literature. The approach is vigorous with elementary calculus only without the use of special functions to this classical EM problem.

For a uniformly charged flat surface, the electric field at a given point in the direction perpendicular to the surface is only dependent on the solid angle subtended by the area to the point. The result is surprisingly simple and appealing.

## Acknowledgements

The author wants to thank many colleagues in the physics department for useful discussions.